\documentstyle[multicol,aps,prb,epsf]{revtex}

\begin{document}

\draft

\title{
Relationship between spiral and ferromagnetic states in the Hubbard model \\
in the thermodynamic limit 
}

\author{
Ryotaro Arita and Hideo Aoki
}
\address{
Department of Physics, University of Tokyo, Hongo,
Tokyo 113-0033, Japan
}
\date{\today}

\maketitle

\begin{abstract}
We explore how the spiral spin(SP) state, 
a spin singlet known to accompany 
fully-polarized ferromagnetic (F) states in the Hubbard model, is related 
with the F state in the thermodynamic limit using the density 
matrix renormalization group and exact diagonalization.  
We first obtain an indication that when the F state is the ground state
the SP state is also eligible as the ground state in that limit.
We then follow the general argument by Koma and Tasaki 
[J. Stat. Phys. {\bf 76}, 745 (1994)] to find that: 
(i) The SP state possesses a kind of order parameter. 
(ii) Although the SP state does not break the 
SU(2) symmetry in finite systems, it does so in the thermodynamic limit by 
making a linear combination with other states that are degenerate 
in that limit. We also calculate the one-particle spectral function 
and dynamical spin and charge susceptibilities 
for various 1D finite-size lattices. We find that
the excitation spectrum of the SP state and the F state is almost identical.
Our present results suggest that the SP and the F states 
are equivalent in the thermodynamic limit.
These properties may be exploited to 
determine the magnetic phase diagram from finite-size studies.
\end{abstract}

\medskip


\begin{multicols}{2}
\narrowtext

\section{introduction}
The itinerant ferromagnetism has been one of the most 
fundamental problems in condensed matter physics, and 
we are still some way from a full understanding of the problem.
For the ferromagnetism in the Hubbard model, a simplest possible 
model for interacting itinerant electrons, 
there are various studies, which are pioneered by 
Gutzwiller\cite{gutz}, Hubbard\cite{hub}, and Kanamori\cite{kana}.
For the Hubbard model on finite-size systems, 
it recently extended to 
rigorous proofs that spin-independent
Coulomb interaction can indeed result in fully-polarized ferromagnetic
ground states without degeneracy for appropriate conditions and/or 
appropriate models\cite{Naga,Mielkem0,Mielkem1,Mielkem2,Mielke99,Tasakim1,mietasa,Tasaki95,kohno}.
Apart from these, there is a body of numerical studies for finite 
systems, where various authors have shown that finite 
Hubbard models with appropriate band fillings and boundary conditions
have fully-polarized ferromagnetic ground 
states\cite{kohno,Daul,Daul2,Liang,saka}.

On the other hand, for infinite systems, 
the itinerant nature of electrons makes the problem interesting and subtle.
The most important reason is 
the existence of the spiral-spin(SP) state,
a {\it spin-singlet} state that has the spin-spin correlation 
length as large as the sample size\cite{Kusakabe94,Kusakabe,watamiya}, 
which is known to accompany the fully-polarized ferromagnetic(F) state 
in various finite-size itinerant models.  
By accompanying we mean SP and F states are close in energy 
in finite systems, where the ground state changes between them 
when the boundary condition is twisted.  

The SP state does not seem to be an accident of the Hubbard model, 
since SP states are found not only in electron systems 
with strong charge-charge interactions, but also 
in those with spin-dependent interactions such as 
double-exchange\cite{Kubo} or Kondo lattice\cite{Sigrist} models.
So we believe that it is a general feature of strongly 
correlated systems for the F state to be accompanied by the SP state.
It is rather surprising, since the latter has the 
total $S=0$ while the former has $S=S_{\rm max}$.  
Thus it is desirable to resolve this question to
understand the ferromagnetism in itinerant systems 
in the thermodynamic limit.
Thus the problem we have to clarify is the relation 
between the SP and F states in the thermodynamic limit (do those states 
degenerate or coalesce, etc), which we address in this paper.   

First, we show that the SP state is eligible as the ground state 
(i.e., degenerate with the F state 
in energy in the thermodynamic limit) 
when the F state is the ground state.  We check this for 
one-dimensional (1D) $t$-$t'$ Hubbard model with 
infinitely large coulomb interaction taken as a typical example 
and calculate the energies of the SP and F states as a function 
of the inverse system size with the 
density matrix renormalization group (DMRG) method. 

Next, we  discuss whether
the symmetry of the SP state is broken in the thermodynamic limit,
following the arguments by Koma and Tasaki\cite{KomaTasaki}
on symmetry breaking and finite-size effects in
quantum many-body systems on a lattice.
We show that (i) the SP state possesses a kind of 
order parameter, (ii) although the state does not 
break the global SU(2) symmetry in the finite system, 
it does so in the thermodynamic limit by 
making a linear combination of other states that are degenerate 
in that limit.  

We have studied excitation spectrum of the SP state by
calculating one-particle spectral function,
dynamical spin and charge susceptibilities for the SP state
in various finite-size models, including 
1D $t$-$t'$ Hubbard model, Tasaki's flat band model, and 
Hubbard ladder. We have found that the excitation spectrum of
these two states in the finite size system are almost identical.

The present results suggest that
we should regard the SP state as identical with the F state 
although they have the total spins in the opposite extremes 
($S=0$ for the SP state, $S=S_{\rm max}$ for F)
in the finite size system.  
These properties can be exploited to 
determine the magnetic phase diagram from the
calculation of the total spin of the
finite-size system in an unambiguous fashion.  
Otherwise we have to worry about 
not only the total spin, but the spin-spin correlation to 
identify SP states in determining the phase diagram.

The paper is organized as follows.
In Sec. II, we first confirm that the SP and F states have 
degenerate energy for $L\rightarrow \infty (L$: sample size) 
in that the energy per site is identical between them.  
DMRG is used to calculate the difference in the total energy, 
$\Delta E \equiv E_{\rm SP}-E_{\rm F}$, between the SP state
and the F state as a function of system size $L$ 
for the $t$-$t'$ Hubbard model with $U=\infty$ as a typical example.  
The extrapolate $\Delta E(L\rightarrow \infty)$ 
is always finite (or zero depending on the boundary condition even 
in the thermodynamic limit), i.e., $\Delta E$ per site does indeed vanish.  
It implies that when the F state is the ground state,
the SP state is also eligible as the ground state if
these states are translation invariant.
Then we move on to the discussion on symmetry breaking of the SP state
along the Koma-Tasaki's argument.
In Sec. III, we calculate the single-particle spectral function 
and dynamical spin and charge susceptibilities for 
various models. Our results suggest that the SP state and the 
F state are identical state in the thermodynamic limit.  

In Sec. IV, we take an appropriate combination of the
boundary condition and the number of electrons to
make the SP state higher in energy than the SF state, and 
the phase diagram for the one-dimensional $t$-$t'$ 
Hubbard model, Tasaki model, and 2-leg Hubbard ladder model is 
obtained with exact diagonalization of small systems.
We have found that the phase boundaries rapidly converge to those 
for larger systems obtained with DMRG. 
We may expect that it can be exploited to the 
determination of the phase diagram of 2D or 
higher dimensional systems.
A summary of the present study is given in Sec. VI.

\section{Discussion on the thermodynamic limit}
\subsection{The energy difference between the SP state and the
F state in the thermodynamic limit}
1D $t$-$t'$ Hubbard model is one of the
simplest models that is thought to exhibit ferromagnetism for
sufficiently strong electron-electron interaction.
M{\"u}ller-Hartmann suggested that in the low-density limit
the ground state of $t$-$t'$ Hubbard model is ferromagnetic\cite{muller}.
Daul and Noack\cite{Daul,Daul2} have carried out highly accurate
DMRG calculation to conclude that there is an extensive ferromagnetic
phase in the phase diagram.
Since the $t$-$t'$ Hubbard model is numerically tractable, 
we study the energy difference of the SP and F states 
by taking this model with $U=\infty$ as an example.

The $t$-$t'$ Hubbard Hamiltonian is given by 
\begin{eqnarray*}
{\cal H}&=&-t\sum_{i=1}^{L}\sum_{\sigma}
(c_{i \sigma}^{\dagger}c_{i+1,\sigma}+{\rm H.c.})\\
&&+t'\sum_{i=1}^{L}\sum_{\sigma}
(c_{i \sigma}^{\dagger}c_{i+2,\sigma}+{\rm H.c.})
+U\sum_{i=1}^{L}n_{i \uparrow}n_{i \downarrow},
\end{eqnarray*}
where $c^{\dagger}_{i \sigma}$ creates 
an electron at site $i$ with spin $\sigma(=\uparrow,\downarrow)$, 
$t$ is the nearest-neighbor hopping, and 
$t'$ the next nearest-neighbor hopping, $U$ the Hubbard repulsion, and 
$n_{i \sigma}\equiv c^{\dagger}_{i \sigma}c_{i \sigma}$.
In this section we take $t=1$, $t'=0.2$ and $U=\infty$. 

As we can check by a simple exact diagonalization,
the ground state of the system has always 
$S_{\rm tot}=0$ for any value of $U$ at least for system 
sizes up to 12 sites 
when a periodic boundary condition(PBC) is adopted.  
This is to be contrasted with the phase diagram obtained by Daul
and Noack with an open boundary condition(OBC), 
in which a wide ferromagnetic region is found for 
$t'=0.2$ and large $U$.  

Let us first confirm that the SP and F states are 
degenerate in the sense that their difference in 
the total energy behaves as
\[
\Delta E \equiv E_{\rm singlet}-E_{\rm F} \rightarrow \,{\rm finite,} 
\]
i.e., the energy per site vanishes, 
\[
\Delta E /L \rightarrow 0, 
\]
where $E_{\rm singlet}$ is the lowest energy 
within the $S_{\rm tot}=0$ sector, namely the energy of the SP state.  

We take OBC, because the DMRG becomes most accurate for this condition. 
As Daul and Noack showed\cite{Daul,Daul2},
the ground state of $t$-$t'$ Hubbard model 
is ferromagnetic for $t'=0.2$ and $U\rightarrow \infty$ in OBC at least
up to 50 sites.
To estimate $\Delta E$, we must calculate the energy of the 
ground state of the $S_{\rm tot}=0$ sector 
which has higher energy than the ferromagnetic ground state.
Therefore, we add the term $\lambda {\bf S}_{\rm tot}^{2}$
to the original Hamiltonian (\ref{eqnttj}) to selectively 
shift the states with
higher total spin to higher energies by turning on $\lambda >0$, 
while conserving the SU(2) symmetry.
Here we set $\lambda =1$.
In Fig. \ref{DMRGspin}, we show the spin correlation function,
$\langle S_i^z S_j^z \rangle$ ($i=L/2, j=L/2+1, \cdots, L-3$)
of the spin singlet state
for $L=32$ sites, $n=0.75$ as an example.
We can see that the spin correlation wave length is as 
large as the system size.
Namely, the spin singlet state is the SP state.

Since DMRG is a variational procedure, the energy of the
SP state is an upper bound, while 
the energy of the F state is calculated exactly because the F state does
not feel the on-site coulomb interaction.
Hence we overestimate $\Delta E$.
To minimize this overestimation, we must calculate the
energy of the SP state as accurately as possible.
This is the reason why we have set $U=\infty$ and $t'=0.2$.
Namely, the exclusion of 
double occupancies reduces the Hilbert space drastically, while 
a small value of $|t'|$ reduces the truncation error.

Using the finite-size algorithm in DMRG, we calculated up to 
$L=48$ for the density of electrons $n=0.5$, 
sweeping the system about 20 times to improve the wave function. 
We store the density matrix at each step to construct good initial vector
for each super-block diagonalization.
We have kept up to $m=500$ states per block at each step, where 
the convergence is checked by comparing the results for $m=300,400,500$.  
The truncation error is smaller than $10^{-5}$, which 
is small enough to enable us to extrapolate 
$\Delta E(L\rightarrow \infty)$.

In Fig. \ref{DMRGene}(solid line), we show the results for 
$\Delta E \equiv E_{\rm SP}-E_{\rm F}$ 
as a function of inverse system size $1/L$
for electron density $n=0.5$.  
We can see that all the points for $L>24$ fall upon 
a linear dependence on $1/L$, from which 
we can extract $\Delta E(L\rightarrow \infty)$.
For all the densities studied, $\Delta E(L\rightarrow \infty)$ 
indeed remains finite ($\sim 10^{-3}$).

In order to check whether the result is not an accident for $U=\infty$, 
we can introduce the effect of large but finite $U$ 
as an (antiferromagnetic) exchange interaction $J>0$,
\begin{equation}
J\sum_{i=1} P_G^{-1}\left( {\bf S_{i}} \cdot {\bf S_{i+1}}
-\frac{1}{4}n_{i}n_{i+1} \right)P_G,
\label{eqnttj}
\end{equation}
added to the $U=\infty$ $t$-$t'$ Hubbard model.  Here 
$S_{i}\equiv \frac{1}{2}\sum_{\beta \gamma}c_{i \beta}^{\dagger}
\vec{\sigma}_{\beta \gamma}c_{i \gamma}$ is the spin operator, where 
$\vec{\sigma}$ is the Pauli matrices and 
$P_G$ denotes the Gutzwiller projection operator.  
In Fig. \ref{DMRGene} the results for 
$J=\pm 0.2$ are superposed.
We can see that $\Delta E$ does not change drastically
even if we introduce the effect of finite $U$.

This kind of $\Delta E$ may depend on the boundary condition, 
which can be indeed the case with finite systems.  For instance, 
Kusakabe and Aoki\cite{Kusakabe} have shown for a finite 
two-dimensional Hubbard model that Nagaoka's ferromagnetic 
state (one hole in the half-filled band with $U=\infty$) 
alternate with an SP state as the boundary condition 
is changed from periodic to anti-periodic.  
In other words a level crossing takes place 
between the two states as an Aharonov-Bohm 
magnetic flux is introduced to twist the boundary condition.  
Such a boundary effect can affect the `energy gap' 
even in the thermodynamic limit in general.

Thus we compare in Fig. \ref{boundary} 
the ordinary OBC (the same as Fig. \ref{DMRGene}, solid line) 
and the boundary condition in which 
we further turn off the nearest-neighbor transfers ($t$) at either end 
of the system in the $t$-$t'$ model 
for $\Delta E(L\rightarrow \infty)$ at $n=0.5$.  
We can see that $\Delta E(L\rightarrow \infty)$ becomes zero
(negligibly small) in the latter case.  
Therefore, we can not exclude a possibility for which
the SP state has a lower energy than the F state
in some appropriate boundary condition.
In other words, the problem which state has a lower energy
is a very subtle problem.

\subsection{The definition of the ground state}
As we have seen, it is meaningless in the thermodynamic limit to 
identify the ground state by studying 
small `finite' differences in the total energy between the 
candidates, so that a totally different point of view is required.
The mathematical definition of the ground state in the
thermodynamic limit is reviewed in Koma and Tasaki's 
article\cite{KomaTasaki} and here we follow them.
For simplicity, we focus on the case of PBC.

We first recapitulate the definition.  
Let $A$ be an arbitrary local operator that acts on a finite number
of sites. We define $\rho$ as 
\begin{eqnarray*}
\rho (A) \equiv \lim_{\Lambda \uparrow Z^d} \rho_{\Lambda}(A)
\end{eqnarray*}
for each $A$, 
and we call $\rho$ as `state' in the thermodynamic limit.
Here $\Lambda$ is a finite ($L\times L\times \cdots L\times L$) 
$d$-dimensional hypercubic lattice while 
$Z^d$ is an infinite $d$-dimensional lattice, 
and we have defined 
\begin{eqnarray*}
\rho_{\Lambda}(\cdots)\equiv
{\rm Tr}_{{\rm H}_\Lambda}[(\cdots)\tilde{\rho}_{\Lambda}],
\end{eqnarray*}
where the trace is taken over the Hilbert space 
${{\rm H}_\Lambda}$ spanned over $\Lambda$,
and $\tilde{\rho}_{\Lambda}$ is an arbitrary density 
matrix on ${{\rm H}_\Lambda}$.
We also assume a Hamiltonian ${\cal H}_\Lambda$ that is local 
as defined by 
\[
{\cal H}_\Lambda=\sum_{x \in \Lambda} h_x.
\]
where $h_x$ is a local component such as 
$h_x = \sum_{\sigma}(c_{x \sigma}^{\dagger}c_{x+1,\sigma}+{\rm H.c.})
+Un_{x \uparrow}n_{x \downarrow}$ for the Hubbard model.
We then define the ground-state energy density $\varepsilon_0$ as
\[
\varepsilon_0 \equiv \lim_{\Lambda \uparrow Z^d} 
\inf_{{\Phi_\Lambda \in {\rm H}_\Lambda},\|\Phi\|=1}
\frac{1}{|\Lambda|}\langle
\Phi_\Lambda |{\cal H}_\Lambda |\Phi_\Lambda \rangle,
\]
where $\Phi$ is a state in ${\cal H}_\Lambda$ and $|\Lambda|$ is 
the norm of $\Lambda$.
A state $\omega$ in the infinite-volume limit is said to be a ground state
if it satisfies
\[
\omega(h_x)=\varepsilon_0
\]
for arbitrary $x$.

In the present context, we have seen in the previous section that 
the numerical results for the Hubbard model 
imply that the energy per site
of the SP state and the F state is identical
in the thermodynamic limit for any boundary condition, 
especially PBC.
In this boundary condition the system has 
a translational invariance, so that 
an SP state in a finite system should have
\[
T(x)|{\rm SP}\rangle_{\Lambda} = \exp(i\gamma)
|{\rm SP} \rangle _{\Lambda}
\]
under the translation $T(x)$ which translates the state by $x$.  
We can expect that the invariance holds in the thermodynamic limit as 
well, so that
\begin{eqnarray*}
\langle {\rm SP}|h_{{x_0}+x}|{\rm SP}\rangle
&=&\langle {\rm SP}|T(x) h_{{x_0}} T^{-1}(x)|{\rm SP}\rangle \\
&=&\langle {\rm SP}|h_{{x_0}}|{\rm SP}\rangle\\
&=&\varepsilon_0,
\end{eqnarray*}
and we can conclude that the SP state in PBC is also 
the ground state in the thermodynamic limit.

\subsection{Symmetry breaking in the spiral spin state}
One important question is: does the SP state, 
despite its being spin singlet, accompany a symmetry
breaking in the thermodynamic limit?
For simplicity, we consider the 1D case. From the numerical results 
for the spin-spin correlation in finite systems, 
a natural quantity that is expected to become nonzero (in PBC) is
\begin{equation}
\frac{1}{L^2}\sum_{i,j\in \Lambda}\cos(\theta_i -\theta_j)
\langle {\bf S}_i \cdot {\bf S}_j \rangle_{\rm SP} >0,
\label{cosSS}
\end{equation}
where $\theta_j = {2 \pi j}/{L}$.
We can then introduce an order parameter,
\begin{eqnarray}
O_\Lambda^{\pm}&\equiv&\frac{1}{L}
\sum_{i \in \Lambda}\exp(\mp i\theta_i)S_i^{\pm}
\equiv\frac{1}{L}\sum_{i \in \Lambda}o_i^{\pm}, 
\label{OP}\\
O_\Lambda^{(1)}&=&\frac{1}{2}(O_\Lambda^+ +O_\Lambda^-) \\
               &\equiv& \frac{1}{L}\sum_{i \in \Lambda}
\cos \theta_i S_i^x +\sin \theta_i S_i^y , \\
O_\Lambda^{(2)}&=&\frac{1}{2i}(O_\Lambda^+ -O_\Lambda^-) \\
               &\equiv& \frac{1}{L}\sum_{i \in \Lambda}
-\sin \theta_i S_i^x +\cos \theta_i S_i^y ,\\
O_\Lambda^{(z)} &\equiv& \frac{1}{L}\sum_{i \in \Lambda} S_i^z 
\end{eqnarray}
where $S^{\pm} = S^x \pm iS^y$.  
Three operators $O_\Lambda^{(1)}, O_\Lambda^{(2)}, O_\Lambda^{(z)}$, 
which obey a Lie algebra, 
satisfy the assumption required for Theorem 2.4 in 
Ref. \onlinecite{KomaTasaki}.  
The expectation values of the squared 
$O_\Lambda^{(1)}$ is
\begin{eqnarray*}
_\Lambda\langle {\rm SP} | O_\Lambda^{{(1)}2} | {\rm SP}
\rangle _\Lambda
&=&\frac{2}{3 L^2} \sum_{i,j\in \Lambda} 
\cos(\theta_i -\theta_j)
\langle {\bf S}_i \cdot {\bf S}_j \rangle_{\rm SP}\\
&&+\frac{1}{2 i L^2}
\sum_{i,j\in \Lambda}\sin(\theta_i-\theta_j)\langle S_i^{+} S_j^{-}
\rangle_{\rm SP}.
\end{eqnarray*}
We can always take 
$|{\rm SP} \rangle_\Lambda$ to be real in PBC 
since ${\cal H}_\Lambda$ is real.  
$S_i^\pm$ can also be represented as 
a real matrix, so that $\langle S_i^{+} S_j^{-} \rangle_{\rm SP}$
is a real number, and we have 
$\langle S_i^{+} S_j^{-} \rangle_{\rm SP}
=\langle S_j^{+} S_i^{-} \rangle_{\rm SP}$ so that the last line above 
vanishes.  From the inequality eq.(\ref{cosSS}) we have
\begin{eqnarray*}
_\Lambda\langle {\rm SP} | O_\Lambda^{{(1)}2} | {\rm SP} \rangle 
_\Lambda
&=&_\Lambda\langle {\rm SP} | O_\Lambda^{{(2)}2} | {\rm SP} \rangle 
_\Lambda\\
&=&\frac{2}{3 L^2} \sum_{i,j\in \Lambda} 
\cos(\theta_i -\theta_j)
\langle {\bf S}_i \cdot {\bf S}_j \rangle_{\rm SP}
>0.
\end{eqnarray*}
On the other hand the order parameters, when not squared, satisfy
\begin{eqnarray*}
_\Lambda\langle {\rm SP} | O_\Lambda^{{(1)}} | {\rm SP} \rangle 
_\Lambda
=_\Lambda\langle {\rm SP} | O_\Lambda^{{(2)}} | {\rm SP} \rangle 
_\Lambda
=0.
\end{eqnarray*}
Thus we have
\[
\Psi_{\Lambda}^{M}\equiv \frac{(O_{\Lambda}^+)^M 
|{\rm SP}\rangle_\Lambda}
{\|(O_{\Lambda}^+)^M |{\rm SP}\rangle_\Lambda \|}
\]
as the `low-lying state' in the Theorem 2.4 
in Ref. \onlinecite{KomaTasaki}, which asserts that 
a symmetry breaking can occur when some low-lying excitations 
whose energies approach that of the state in question 
like $1/L$ mix with it. 
Here $\Psi_{\Lambda}^{M}$ does 
have an energy that approaches to $E_{\rm SP}$ as
\[
|(\Psi_{\Lambda}^{M}, H_{\Lambda} \Psi_{\Lambda}^{M})
-E_{\rm SP} | \leq {\rm const.}\times (\frac{M^2}{L}).
\]
We also have, according to Theorem 2.5\cite{KomaTasaki},
\[
\lim_{k\rightarrow \infty}\lim_{\Lambda\uparrow Z}
(\Xi_{\Lambda}^k, O_{\Lambda}^{(1)}\Xi_{\Lambda}^k)
={\rm const.} >0
\]
where, 
\[
\Xi_{\Lambda}^{k} \equiv \frac{1}{\sqrt{2k+1}}
\left[|{\rm SP}\rangle_\Lambda + \sum_{M=1}^{k}
\left( \Psi_{\Lambda}^{M}+\Psi_{\Lambda}^{-M}
\right)\right].
\]
Here, let us introduce the translation operator $T(x)$, which
satisfies $T(x) c_{x_0}^\dagger T^{-1}(x)=c_{x_0+x}^\dagger$.
Since 
\begin{eqnarray*}
T(x)O_\Lambda^+T^{-1}(x)&=&\frac{1}{L}\sum_{x_0\in \Lambda} 
\exp(-i \theta_{x_0})T(x)S_{x_0}^+T^{-1}(x) \\
&=&\frac{1}{L}\sum_{x_0\in \Lambda} \exp (-i\theta_{x_0}) S_{x+x_0}^+ \\
&=&\exp (i \theta_x)O_\Lambda^+,
\end{eqnarray*}
$T(x)\Psi_{\Lambda}^M=\exp(iM\theta_x)\exp(i\gamma)\Psi_{\Lambda}^M$,
where $\gamma$ is a constant.
Thus, for $o_i^+ \equiv {\rm e}^{-i\theta_i}S_i^+$ appearing in 
eq. (\ref{OP}),
\begin{eqnarray*}
_\Lambda\langle 
\Psi_{\Lambda}^{M}| o_{x+x_0}^{+} |\Psi_{\Lambda}^{M-1}
\rangle_\Lambda &=&
_\Lambda\langle \Psi_{\Lambda}^{M}| {\rm e}^{-i\theta_{x+x_0}}S_{x+x_0}^{+} 
|\Psi_{\Lambda}^{M-1}\rangle_\Lambda \\
&=&
_\Lambda\langle \Psi_{\Lambda}^{M}| 
{\rm e}^{-i\theta_{x}}T(x)o_{x_0}^{+} T^{-1}(x)
|\Psi_{\Lambda}^{M-1}\rangle_\Lambda \\
&=&{\rm e}^{( -i\theta_x+iM\theta_x-i(M-1)\theta_x+i\gamma-i\gamma)}\\
&&\times  _\Lambda\langle 
\Psi_{\Lambda}^{M}| o_{x_0}^{+} |\Psi_{\Lambda}^{M-1}\rangle_\Lambda \\
&=&
_\Lambda\langle 
\Psi_{\Lambda}^{M}| o_{x_0}^{+} |\Psi_{\Lambda}^{M-1}\rangle_\Lambda.
\end{eqnarray*}
Therefore, we have
\begin{eqnarray*}
\lim_{k\rightarrow \infty}\lim_{\Lambda\uparrow Z}
\frac{1}{|\Omega|}\sum_{x\in \Omega}
\langle \Xi_{\Lambda}^k |o_x^+| \Xi_{\Lambda}^k \rangle 
&=&\lim_{k\rightarrow \infty}\lim_{\Lambda\uparrow Z}
\frac{1}{|\Omega|}\sum_{x\in \Omega}
\langle \Xi_{\Lambda}^k |o_{x+x_0}^+| \Xi_{\Lambda}^k \rangle \\
&=&{\rm const.}>0,
\end{eqnarray*}
where $\Omega$ is a finite subspace of $\Lambda$ whose size is $|\Omega|$.
Hence,
\begin{eqnarray*}
\lim_{k\rightarrow \infty}\lim_{\Lambda\uparrow Z}
\frac{1}{|\Omega|}\sum_{x\in \Omega}
\langle \Xi_{\Lambda}^k |S_{x}^+| \Xi_{\Lambda}^k \rangle 
={\rm const.} >0
\end{eqnarray*}
Namely, although the SP state itself does not 
break the symmetry, it does so in 
in the thermodynamic limit 
by letting $\{\Psi_{\Lambda}^{M}\}$ mix with it in that limit.
In other words, an arbitrarily large but finite region in 
an SP state has a broken SU(2) with a finite magnetization.

If we use identities 
$S^{-}=\sum_{x_i \in \Lambda}S_i^{-}$, $[O^-,S^-]=0$,
$\langle {\rm SP}|S^- = 0$,
\begin{eqnarray*}
&&\langle \Psi _{\Lambda}^M | {\rm F}; S^z=M \rangle \\
&&=_\Lambda\langle {\rm SP}| (O_{\Lambda}^-)^M 
(S^-)^{S_{\rm max}-M} |{\rm F}; S^z=S_{\rm max} \rangle _\Lambda \\
&&= 0.
\end{eqnarray*}
Thus both of the SP state and $\{\Psi_{\Lambda}^{M}\}$ 
that mixes with it in the thermodynamic
limit turn out to be orthogonal to the F state
for finite systems.  

Finally, let us discuss whether $\Xi$ is the ground state in the
thermodynamic limit by checking whether 
$\langle \Xi|h_x|\Xi\rangle=\varepsilon_0$ holds 
for arbitrary $x$. 
Since 
\begin{eqnarray*}
&&_\Lambda\langle {\rm SP}|(O_{\Lambda}^-)^M
h_{{x_0}+x} (O_{\Lambda}^+)^M |{\rm SP} \rangle_\Lambda \\
&&=_\Lambda\langle {\rm SP}|(O_{\Lambda}^-)^M
T(x)h_{{x_0}}T^{-1}(x) (O_{\Lambda}^+)^M |{\rm SP} \rangle_\Lambda \\
&&=_\Lambda\langle {\rm SP}|(O_{\Lambda}^-)^M
h_{{x_0}}(O_{\Lambda}^+)^M |{\rm SP} \rangle_\Lambda\\
&&=\varepsilon_0,
\end{eqnarray*}
$(\Psi_{\Lambda}^M, h_x \Psi_{\Lambda}^M)=\varepsilon_0$ does indeed hold 
for arbitrary $x$ in the limit of $L\rightarrow \infty$.
This means $\Xi$ is the ground state.

\section{The excitation spectrum of the spiral spin state}
Having shed light in the context of Koma-Tasaki's argument, 
we may now conjecture that the SP state and the F state
are identical in the thermodynamic limit.
Thus let us have a closer look at the SP state as compared with the F state 
in three one-dimensional models, i.e., the $t$-$t'$ Hubbard model, 
Tasaki's flat band model, and the two-leg Hubbard ladder.
To determine whether the SP and F states
are identical in the thermodynamic limit,
we must show that the relation
$\omega_{\rm SP}(A)=\omega_{\rm SF}(A)$ holds 
for arbitrary local operator $A$.
Although it is impossible to confirm this 
for all the possible $A$, if the F state and the SP state are
identical in that limit, the excitation spectrum 
of these state in finite size system are expected to
be almost identical. To confirm this,
we look specifically at the single-particle spectral function 
and the dynamical spin and charge correlation functions of the SP state.
Indeed we will show that since the nature of the SP state 
in finite systems is almost identical to 
that of the F state, we cannot distinguish the SP state from the
F state by looking at these quantities for sufficiently large system.

\subsection{$t$-$t'$ Hubbard model}

In the $t$-$t'$ Hubbard model where the electron transfer 
extends to next nearest neighbors ($t'$), 
the ground state is $S_{\rm tot}=0$ for any value of $U$ in PBC 
at least for sizes up to 12 sites as mentioned in the previous section.  
Let us first look at the spin-spin correlation function, 
$\langle \Phi_G | S_i^z S_j^z | \Phi_G \rangle$, 
to confirm that the $S_{\rm tot}=0$ ground state 
for large enough $U\geq U_C$ is the SP state.
We show in Fig. \ref{spinspinco} the result for 
10 electrons in a 12-site ring (two holes) with $U=20$ or 40.
For $U=20$, a short-range 
antiferromagnetic spin-spin correlation is observed 
for $\langle S_i S_{i+1}\rangle <0$.
By contrast, for $U=40$ (or larger, which falls upon 
the ferromagnetic region in the phase diagram; see 
Fig. \ref{apbcphasett02} below), 
the spin-spin correlation has a wave length as large as the system size,
which is the same behavior of the lowest $S_{\rm tot}=0$ state in OBC 
(Fig. \ref{DMRGspin}).

A more detailed nature the states is encapsulated in 
the single-particle spectral function given by
\begin{eqnarray}
A^{\pm}(k,\omega)=\frac{1}{\pi}{\rm Im}
\langle \Phi_G | \gamma^{\mp}_{k \sigma}
\frac{1}{\omega\pm(E_0-{\cal H})-i0}
\gamma^{\pm}_{k \sigma}|
\Phi_G \rangle,
\label{akwdef}
\end{eqnarray}
where $A^{+}(A^{-})$ denote the electron addition (removal) spectrum 
with $\gamma^{+}_{k \sigma}\equiv c^{\dagger}_{k \sigma}$, 
$\gamma^{-}_{k \sigma}\equiv (\gamma^{+}_{k \sigma})^{\dagger}$, 
$| \Phi_G \rangle$ and $E_0$ the ground state and its energy,
respectively. 
We have numerically calculated this quantity.  
In Fig. \ref{akw1}, we show the result 
for the parameter values adopted in Fig.\ref{spinspinco} with $U=40$. 
We can see that the Luttinger relation,
$k_F=\pi n/2$ (with $n=10/12=0.83$ here), 
for the spin unpolarized electrons  
is violated in favor of $k_F=\pi n$,
which would be expected for fully spin-polarized fermions. 
This behavior in the single particle
spectral function is consistent with the nearly ferromagnetic
nature of the SP state. 
Such a behavior was 
first found for the $t$-$t'$-$J$ model by Eder and Ohta\cite{EderOhta}, 
and then given an interpretation subsequently\cite{Arita}.

The $k$-dependence of the spectral function may be analyzed 
by evoking the argument 
by Doucot and Wen(DW)\cite{DoucotWen}, who studied
the infinite-$U$ Hubbard model on a two-dimensional 
square lattice with two holes to 
find a trial state that gives an energy lower 
by $1/L^2$ than Nagaoka's ferromagnetic state.  
The key idea of DW is based on the following intuition.  
Holes behave as free fermions for on-site interactions 
when the background spin state is ferromagnetic (or nearly so). 
Assume that fully 
polarized electrons take an open-shell configuration 
(in which there is a degeneracy in the free-electron configurations) 
in $k$-space in PBC.  
If the background spin state is changed from F to the SP spin texture, 
a fictitious gauge field corresponding to half flux quantum is generated. 
This shifts the $k$-points by half the $k$-point spacing, which 
lowers the kinetic energy because the polarized electrons now take 
a closed-shell configuration.  

It is a nontrivial question whether 
the DW's trial state may be applied to reproduce the 
spinless-fermion-like behavior in $A(k,\omega)$ found here for 
the 1D system.   
Let us assume, for simplicity, that the holes hop in a rigid spin 
background, although in the original paper DW considered
the holes dressed by spin waves.  
The energy of the hole in the $t$-$t'$ Hubbard model
is then given as 
\begin{eqnarray}
\varepsilon&=&-2t
\cos\left(k \pm \frac{\pi}{L}\right) 
+2 t' 
\cos 2\left(k \pm \frac{\pi}{L}\right),
\label{energy1}
\end{eqnarray}
where $k(=2\pi N/L, 0 \le N \le L, N$ is an integer) is wave number
and $\pm$ corresponds 
to the sign of the fictitious flux.  
We can see that the fitting in Fig. \ref{akw1} is remarkably good. 
To be precise, the transfer integrals in the original DW 
are taken to be effective ones 
($t \rightarrow t \cos (\pi/L), t' \rightarrow t' \cos (2\pi/L)$) 
to take care of the reduction in the transfer between 
slightly twisted spins, but the present result 
is better fitted with these reduction factors omitted (dashed curves).   

In the addition spectrum an almost dispersionless band
of low intensity peaks is seen (at around $E=2$ in Fig. \ref{akw1}). 
If the ground state were the fully polarized F state, 
then such a band is expected, since 
we can add an opposite spin at any $k$-point.  
A remnant of such a band 
again suggests the nearly ferromagnetic nature of the SP state. 

A next question is whether the SP state persists 
for more than two holes.  
In Fig. \ref{spinspinco2}, we show the spin-spin correlation
for 6 electrons on a 12-site lattice (six holes, 
or $n=0.5$; quarter filled) in PBC for $U=4,6$.  
We can see that for $U=6$, the spin-spin correlation is again spiral.
To investigate whether DW's picture is valid for such a high hole doping, 
we have calculated the single-particle spectrum for $U=6$ 
in Fig. \ref{akw2}.  
The spectrum for the SP state is fitted well by the energy dispersion
defined by the eq.(\ref{energy1}), again $t$ and $t'$ not reduced, 
even though the hole concentration is as large as $n_h=0.5$.
This is surprising, since the assumption that
holes are nearly free would be valid only for small enough doping.   

We next question whether the SP state is connected adiabatically to 
the antiferromagnetic state as we decrease the Hubbard $U$.  
We show the ground-state energy as a function of $1/U$ for 
6 electrons on a 12-site ring in Fig. \ref{level}.
We can clearly identify a level crossing appearing as a cusp 
around $U \simeq 5$, 
which indicates that a transition (rather than an adiabatic 
connection) occurs within the $S=0$ 
space from the antiferromagnetic
phase to the SP phase.  

We now turn to the dynamical spin and charge correlation
functions 
\begin{eqnarray*}
N(k,\omega)&=&\frac{1}{\pi}{\rm Im}
\langle \Phi_G | N_{k}
\frac{1}{\omega+(E_0-{\cal H})-i0}
N_{k}|
\Phi_G \rangle ,\\
S(k,\omega)&=&\sum_\alpha\frac{1}{\pi}{\rm Im}
\langle \Phi_G | S_{k}^\alpha
\frac{1}{\omega+(E_0-{\cal H})-i0}
S_{k}^\alpha|
\Phi_G \rangle ,\\
\end{eqnarray*}
where $N_k$  and $S_k^\alpha$ are  Fourier transform of 
$(n_i-n)$ and $S_i^\alpha(\alpha=x,y,z)$, respectively.
In Fig. \ref{dynamical-s-c}, we show the results
for the F state in APBC and
the SP state in PBC for 10 electrons in 12 sites
with $U=40$. 
We can see that the behavior of these functions for the SP state
is surprisingly similar to those of the F state even for 
$L=12$. We can expect that
they will become identical in the thermodynamic limit.

\subsection{One-dimensional Tasaki model}
Tasaki\cite{Tasakim1} proposed a Hubbard model on a 
special class of lattice structures for which the lowest
energy band is dispersionless (flat).
He proved rigorously that the ground state is ferromagnetic
for arbitrary interaction strength ($0<U<\infty$) 
when the flat band is half-filled.

To clarify whether metallic ferromagnetism can be realized in
models having a non-half-filled, nearly-flat band,
Sakamoto and Kubo\cite{saka} investigated with DMRG the
Hubbard model on a chain of triangles, 
which may be thought of as a realization of Tasaki's model in 1D.  
They obtained results which suggest that
the system exhibits metallic, fully-polarized ferromagnetism for sufficiently
strong interaction.
On the other hand, Watanabe and Miyashita\cite{watamiya} found 
that an SP state is the ground state 
at least for system sizes up to $L=12$ 
for appropriate conditions,
i.e., sufficiently large $U$, $n>0.5$, even number of electrons and PBC.

The Hamiltonian is given by
\begin{eqnarray*}
{\cal H}&=&
t\sum_{i=1}^{L/2}\sum_\sigma(c_{2i,\sigma}^{\dagger}
c_{2(i+1),\sigma}+{\rm H.c.})\\
&&+\alpha t \sum_{i=1}^{L/2}\sum_\sigma
[(c^{\dagger}_{2i-1,\sigma}+c^{\dagger}_{2i+1,\sigma})c_{2i,\sigma}+{\rm H.c.}]\\
&&+\beta t \sum_{i=1}^{L/2}\sum_\sigma n_{2i-1,\sigma}
+U\sum_{i=1}^{L}n_{i\uparrow}n_{i\downarrow},
\end{eqnarray*}
where 
$t$ is the transfer for the bottom bond in a triangle, 
$\alpha t$ for the bonds connecting bottom and top sites and
$\beta t$ the site energy of a top site.  
Since a unit cell contains two sites, 
the single-electron spectrum consists of two bands 
with the dispersion relation, up to a constant,
\begin{eqnarray*}
\varepsilon^{\pm}&=&\frac{t}{2}
\left[ 2 \cos k \pm 
\sqrt{4\cos^2 k+(8\alpha^2-4\beta )\cos k+ \beta^2 +8\alpha^2}
\right].
\end{eqnarray*}
The lower band is then flat when $\beta=\alpha^2-2$ is satisfied.  
Hereafter in this subsection,
we take $t=1$, $\alpha=2$, $\beta =0$ 
(for which the lower band is dispersive)
and $L$ even.

While Watanabe and Miyashita have shown that SP states 
appear for more than quarter-filled bands ($0.5<n<1.0$), 
we find here that the SP state is the ground state of $S_{\rm tot}=0$
sector for sufficiently large $U$ 
in the region $n\leq 0.5$ as well at least for $L\leq 12$.
In Fig. \ref{spincortasaki}, 
we show the spin correlation for 6 electrons on a 12-site
lattice in PBC for $U=4$ and $U=6$. 
We can see that for the case of $U=6$ the spin correlation length
is indeed as large as the system size.
We have also calculated the single-particle spectral function, 
eq.(\ref{akwdef}), 
where we take 
$\gamma^{+}_{i,\sigma}=c^{\dagger}_{2i, \sigma}$
or $\gamma^{+}_{i,\sigma}=c^{\dagger}_{2i-1, \sigma}$.  
If we combine them, they should contain 
the contributions from the two bands.   
Here $\gamma^{+}_{i\sigma}$ is a Fourier transform of
$\gamma^{+}_{k\sigma}$.
In Fig. \ref{akwtasaki}, we show the result for 6 electrons on 12 site lattice,
where the spectrum is fitted by a two-band extension of eq.(\ref{energy1}),
\begin{eqnarray}
\varepsilon^{\pm}=
 t\cos (k\pm\frac{\pi}{L}) 
\pm t\sqrt{\cos^2 (k\pm\frac{\pi}{L}) +8\cos (k\pm\frac{\pi}{L}) +8}.
\label{energy3}
\end{eqnarray}
Fig. \ref{akwtasaki} shows that the DW picture 
(almost free electrons hopping 
in a twisted spin background) is surprisingly good even for 
the two-band case.

\subsection{Two-leg Hubbard ladder}
Ferromagnetic ground state in 1D has also been obtained for ladder systems, 
where two chains are connected with an inter-chain transfer, $t_{\perp}$.
Liang and Pang\cite{Liang} suggested  that for 
$t_{\perp}/t=1$ and $U=\infty$,
the F state is one of the ground states for $n_h <0.22$, 
while the ground state is spin singlet for $n_h >0.4$.
Kohno\cite{kohno} presented rigorous results for
$U=\infty$ 2-leg Hubbard ladder in the limit of large inter-chain
hopping ($t_{\perp}/t \rightarrow \infty$).
He also studied the case for finite values of
$t_{\perp}/t$ with DMRG to obtain the phase diagram.
His results are consistent with that of Liang and Pang.
These studies assumed OBC.

On the other hand, 
it is intriguing to study whether an SP 
state appears in the ladders as well for PBC, and if so, whether 
$A(k,\omega)$ exhibits a DW-like behavior.
The Hamiltonian is given by
\begin{eqnarray*}
{\cal H}&=&-t\sum_{i=1}^{L}\sum_{\alpha,\sigma}
(c_{i,\alpha \sigma}^{\dagger}c_{i+1,\alpha\sigma}+{\rm H.c.}) \\
&&-t_{\perp}\sum_{i=1}^{L}\sum_{\sigma}
(c_{i,1\sigma}^{\dagger}c_{i,2\sigma}+{\rm H.c.})
+U\sum_{n=1}^{L}\sum_{\alpha}n_{i \alpha \uparrow}n_{i \alpha \downarrow},
\end{eqnarray*}
where $\alpha(=1,2)$ labels the two legs of the ladder.
We set $t=1$, $U=\infty$.

We have calculated the intra-chain 
spin-spin correlation $\langle S_{i \alpha}^z
S_{j \alpha}^z \rangle$ (Fig.\ref{spincorladd}), 
and found that the ground state is
indeed SP. We have also calculated the spectral function, 
eq.(\ref{akwdef}), where we now take 
$\gamma^{+}_{k\sigma}=c^{\dagger}_{k 1\sigma}+c^{\dagger}_{k 2\sigma}$
(creating an electron in the bonding band) or 
$\gamma^{+}_{k\sigma}=c^{\dagger}_{k 1\sigma}
-c^{\dagger}_{k 2\sigma}$ (antibonding).
In Fig.\ref{akwlad}, we show the result for two holes (14 electrons on
a $8 \times 2$ ladder).
We fit the spectrum by the dispersion,
\begin{eqnarray}
\varepsilon=-2t
\cos\left(k \pm \frac{\pi}{L}\right) \pm t.
\label{energy2}
\end{eqnarray}
Again the picture of two holes hopping 
in a twisted spin background gives an accurate description.

\section{Magnetic phase diagram from finite-size studies}
Our results in the previous sections
suggest that we should regard
an SP state  as being ferromagnetic rather than non-magnetic
even though $S_{\rm tot}=0$.  
So we finally come to the problem of 
how to determine the magnetic phase diagram from finite-size 
studies in the light of the SP state.  
As we have stressed, the ground state of finite systems 
in a certain boundary condition always has
$S_{\rm tot}=0$ no matter how $U$ is strong 
in some one-dimensional models as exemplified by the 
$t$-$t'$ Hubbard model and the Hubbard ladder.
Even when a ferromagnetic state appears for finite systems, 
the magnetic region in the phase diagram can shrink for some 
system size as shown by Sakamoto and Kubo\cite{saka} 
for the 1D Tasaki's model.   
This is due to the appearance of the SP state, which are encountered 
if we assume PBC for even number of electrons, or more generally,
if we assume a boundary condition for which the fully-polarized electrons
take an open-shell configuration in the ground state.  

Since the SP state may be regarded as ferromagnetic as 
elaborated in previous sections, 
we have then to distinguish the SP state from nonmagnetic states 
by calculating the spin correlation, etc. 
Here we propose that this difficulty can be readily overcome 
by taking an appropriate
combination of the boundary condition and the 
electron number for which the SP
state is excluded from the ground state --- 
this enables us to obtain a reliable magnetic 
phase diagram within finite-size studies by simply looking at 
$S_{\rm tot}$ without worrying about the existence 
of $S_{\rm tot}=0$ ferromagnetic-like states.
We stress here that this procedure is allowed because the SP state 
may be regarded as ferromagnetic: if the SP state were a 
distinct state, then the exclusion of SP states would lead to 
a missed phase transition. 

Here we illustrate this with an appropriate boundary condition
that makes the F state take a closed-shell configuration.
We present the phase diagram thus obtained 
for the models employed in section III, i.e., 1D $t$-$t'$ Hubbard model,
Tasaki's model, and 2-leg Hubbard ladder.
In all cases, the results accurately coincide with those obtained by
DMRG calculation for much larger systems with OBC.
In other words, the result converges to the thermodynamic 
limit rapidly when we concentrate on the F states.
We may expect that this method should be applicable to the 
determination of the phase diagram of 2D or 
higher dimensional systems.

\subsection{$t$-$t'$ Hubbard model}
We first observe that, in the one-electron energy band of 
the $t$-$t'$ Hubbard model,
\begin{eqnarray*}
\varepsilon (k)=-2 t \cos k +2 t' \cos(2 k),
\end{eqnarray*}
the band minimum at $k=0$ splits
into double minima for $t' >0.25$ as $t'$ is increased.  
Hereafter in this section, we set $t=1$.

We first look at the case of $t'=0.2<0.25$.
The band has a single minimum, so that 
an even number of fully-polarized electrons
take a closed-shell configuration for all densities $n<1$
if we assume APBC to put the $k$-points symmetrically about $k=0$.  
In Fig. \ref{apbcphasett02},
we show the exact diagonalization results for
4, 6, 8, or 10 electrons in a 12-site system ($0.33\leq n\leq 0.83$), and 
4, 6, or 8 electrons in a 10-site system.  
We have also plotted the DMRG result for the phase boundary
obtained by Daul and Noack for a 50-site system with OBC\cite{Daul,Daul2}.
We can immediately see that the results obtained with APBC for fairly 
small systems is close to that obtained for
larger ones.

If we move on to the case of a larger $t'=0.8$,
an even number of fully-polarized electrons only 
take a closed-shell configuration for APBC 
when the Fermi energy is higher than $-2t+2t'$.  
When the Fermi energy is lower, the situation depends on 
how the $k$-points are located around the double-minimum 
dispersion in the given boundary condition, 
so we concentrate on the former case.  
In Fig. \ref{apbcphasett08},
we show the results for 8 or 10 electrons in a 12-site system 
and 8 electrons in a 10-site system.
we can see that the result for small systems with APBC 
is very close to the DMRG result for a 50-site system.

\subsection{One-dimensional Tasaki's model}
The phase diagram against $U$ and $\beta$ (that
controls the dispersion of the band)
of the 1D Tasaki's model (a chain 
of triangles) was obtained by
Sakamoto and Kubo\cite{saka} 
for various band fillings ($n=1/2, 1/4, 3/8$)
by means of the DMRG method with PBC. As they mention,
the phase boundary between $S_{\rm tot}=S_{\rm max}$ and 
$S_{\rm tot}=0$ varies strongly with the system size for small systems, 
so that a reliable phase diagram can be obtained only by going to 
sufficiently large ($L=32$) systems.

In Fig. \ref{phaseflat}, we show the phase boundary determined
by exact diagonalization 
for $n=1/2$ and system size as small as 
8 or 12 sites with APBC.  
Here we fix $\alpha = 2$,
so that the lower band becomes flat for $\beta =2$.
Again, the result excellently agrees
with that of 32 sites with PBC. 

\subsection{Two-leg Hubbard ladder model}
As mentioned in Sec. IIIC, Kohno\cite{kohno}
obtained the phase diagram of $U=\infty$ Hubbard ladder
by means of the DMRG method.
According to his results, there is a wide region
of partially ferromagnetic phase in the neighborhood of
paramagnetic phase.

Let us consider whether we can reproduce the phase boundary
between non-magnetic phase and partially polarized phase
by assuming APBC for small systems.
We consider the case of 6 and 8 electrons on a $6\times 2$ lattice
and 4 and 6 electrons on a $4\times 2$ lattice.
We performed calculation for $0.125 \leq t_{\perp}/(t_{\perp}+t) \leq 0.909$,
and found the ferromagnetic-non-magnetic transition 
only for the case of 8 electrons on
a $6\times 2$ lattice, around $t_{\perp}/(t_{\perp}+t) \sim 0.45$.
This is consistent with Kohno's the result\cite{kohno}.

All these results indicate that the phase boundary between the ferromagnetic
and the antiferromagnetic states is reasonably close to the
thermodynamic limit for small systems when a care is taken 
(i.e., letting fully-polarized electrons 
take closed shell configurations).

\section{Summary}
To summarize, we have studied
the relation between the fully-polarized ferromagnetic state and
the spiral spins state, a
spin-singlet spin state that has a spin correlation length 
as large as the system size, which accompanies 
the fully-polarized ferromagnetic state in a number of 
electron correlation models.
As a typical example, we calculate
the energy of the SP state and the F state
for one-dimensional $U=\infty$ $t$-$t'$ Hubbard model,
It suggests that the SP state is also the ground state
in the thermodynamic limit when the F state is 
the ground state.
Following the argument by Koma and Tasaki\cite{KomaTasaki},
we have also indicated how a symmetry becomes broken in the SP state 
in the thermodynamic limit, where some `low-lying states' become hybridized.

We have then characterized the SP state, an itinerant magnetic state, by 
calculating the one-particle spectral function.   
The result is interpreted in a picture
in which holes move almost freely in a twisted spin configuration.
We have also calculated the dynamical spin and charge 
correlation functions 
to find that their behaviors in the SP state are similar to those of 
the F state.  

From these we have conjectured that we should regard the SP state and
the F state are equivalent in the thermodynamic limit,
even though the SP state (spin-singlet) 
and the F state (fully-polarized) are opposite extremes in 
terms of ${\bf S}_{\rm tot}$. 
We have then shown that the magnetic phase diagram can be 
determined accurately from finite-size studies by 
taking appropriate boundary condition (that depends on 
the number of electrons which are accommodated in 
the shells in the non-interacting case) that pushes up  
the energy of the SP state.  This enables us to 
concentrate on the F states, i.e., to simply look at 
the change in $S_{\rm tot}$.  This is permissible 
since the SP and F states are regarded to be equivalent, 
so that we can concentrate on either of them.  
We have obtained the phase diagram in this way for
1D $t$-$t'$ Hubbard model, Tasaki's model and 2-leg Hubbard
ladder.  
We have found that the phase boundary between the ferromagnetic state
and the non-magnetic state determined in such a way accurately 
coincides with those obtained by the DMRG calculation for 
much larger systems.
Since the method does not depend on the dimensionality, 
we may expect that it should be applicable to the 
determination of the phase diagram in two or 
higher dimensional systems.

\section{Acknowledgments}
We would like to thank Kazuhiko Kuroki, Koichi Kusakabe,
Tohru Koma for illuminating discussions.  
R.A. is supported by Japan Society for Promotion of Science.
Numerical calculation were done on 
FACOM VPP 500/40 at the Supercomputer center, Institute
for Solid State Physics, University of Tokyo.

\begin{figure}
\epsfxsize=7cm
\caption{
The DMRG result
for the spin-spin correlation function for 24 electrons on a 
32-site $t$-$t'$ model with OBC.
}
\label{DMRGspin}
\end{figure}

\begin{figure}
\epsfxsize=7cm
\caption{
The DMRG result for the energy difference between the SP state and the
ferromagnetic state, $\Delta E \equiv E_{\rm SP}-E_{\rm F}$),
in 1D $t$-$t'$ Hubbard model 
as a function of inverse system size $1/L$ for the density 
of electrons $n=0.5$ with the exchange interaction 
$J=0$ (solid lines), $J=0.2$ (dashed lines), or $J=-0.2$ (dotted lines).
}
\label{DMRGene}
\end{figure}

\begin{figure}
\epsfxsize=7cm
\caption{
The energy difference between the SP state and the
ferromagnetic state, $\Delta E \equiv E_{\rm SP}-E_{\rm F}$),
as a function of inverse system size $1/L$ 
for the density of 
electrons $n=0.5$ with the 
ordinary OBC (dashed line) or 
the boundary condition that omits the hopping at the both
ends of the system (solid line).
}
\label{boundary}
\end{figure}

\begin{figure}
\epsfxsize=7cm
\caption{
The spin-spin correlation function for 10 
electrons on a 12-site $t$-$t'$ Hubbard model with $U=40$ 
and the periodic boundary condition.
}
\label{spinspinco}
\end{figure}

\begin{figure}
\epsfxsize=7cm
\caption{
Single particle spectral function for 10 electrons on a 
12-site $t$-$t'$ Hubbard model with $U=40$. 
Solid lines denote the electron-removal spectrum, while 
dotted lines the electron-addition spectrum. 
Dashed curves indicate the energy dispersion defined by
eq. (\protect\ref{energy1}).
}
\label{akw1}
\end{figure}

\begin{figure}
\epsfxsize=7cm
\caption{
A plot similar to Fig. \protect\ref{spinspinco} for 6 electrons 
on a 12-site $t$-$t'$ model with
$U=4$ (solid line) and 6 (dashed line).
}
\label{spinspinco2}
\end{figure}

\begin{figure}
\epsfxsize=7cm
\caption{
A similar plot to Fig. \protect\ref{spinspinco}
for 6 electrons on a 12-site $t$-$t'$ model with $U=6$.
}
\label{akw2}
\end{figure}

\begin{figure}
\epsfxsize=7cm
\caption{
The ground state energy as a function of $1/U$ 
for 6 electrons on a 12-site $t$-$t'$ model.
}
\label{level}
\end{figure}

\begin{figure}
\epsfxsize=6cm
\epsfxsize=6cm
\caption{
The dynamical spin(a) and charge(b) 
correlation functions for the SP state (dashed line) and 
the F state (solid line)
for 10 electrons on a 
12-site 1D $t$-$t'$ Hubbard model with $U=40$.
}
\label{dynamical-s-c}
\end{figure}

\begin{figure}
\epsfxsize=7cm
\caption{
The spin-spin correlation function for 10 electrons on a 
12-site 1D Tasaki's model for $U=4, 6$.
}
\label{spincortasaki}
\end{figure}

\begin{figure}
\epsfxsize=7cm
\caption{
A plot similar to Fig. \protect\ref{akw1}
for 6 electrons on a
12-site 1D Tasaki's model for $U=12$.
We have superposed the results when 
we take $\gamma^{+}_{i,\sigma}=c^{\dagger}_{2i,\sigma}$
and $\gamma^{+}_{i,\sigma}=c^{\dagger}_{2i-1,\sigma}$.  
Dashed curves indicate the dispersion defined by 
eq. (\protect\ref{energy2}).
}
\label{akwtasaki}
\end{figure}

\begin{figure}
\epsfxsize=7cm
\caption{
The spin-spin correlation function for 10 electrons on 
a $8\times$2 $U=\infty$ Hubbard ladder.
}
\label{spincorladd}
\end{figure}

\begin{figure}
\epsfxsize=7cm
\caption{
A similar plot to Fig. \protect\ref{akw1} for 14 electrons on a
$8 \times 2$ ladder.
Dashed curves indicate the dispersion defined by 
eq. (\protect\ref{energy3}).
}
\label{akwlad}
\end{figure}

\begin{figure}
\epsfxsize=7cm
\caption{
Phase diagram of 1D $t$-$t'$ Hubbard model for
$t'=0.2$ obtained by exact diagonalization 
for a 12-site system(diamonds) or 10-site system(circles).  
A DMRG result of 
Daul and Noack\protect\cite{Daul,Daul2} is superposed (a full line).
}
\label{apbcphasett02}
\end{figure}

\begin{figure}
\epsfxsize=7cm
\caption{
A plot similar to Fig. \protect\ref{apbcphasett02}, 
for the $t$-$t'$ Hubbard model with $t'=0.8$.
}
\label{apbcphasett08}
\end{figure}

\begin{figure}
\epsfxsize=7cm
\caption{
Phase diagram against $U$ and $\beta$ 
for 1D Tasaki's model with $n=1/2$
for a 12-site system(diamonds) or 8-site system(circles) with PBC.  
A DMRG result by Sakamoto and Kubo\protect\cite{saka}
are superposed (a solid line).
}
\label{phaseflat}
\end{figure}

\end{multicols}

\begin{references}
\bibitem{gutz}M. C. Gutzwiller, Phys. Rev. Lett. {\bf 10},159 (1963).
\bibitem{hub}J. Hubbard, Proc. R. Soc. London, Ser. A {\bf 276}, 238 (1963).
\bibitem{kana}J. Kanamori, Prog. Theor. Phys. {\bf 30}, 275 (1963).
\bibitem{Naga}Y. Nagaoka, Phys. Rev. {\bf 147}, 392 (1966).
\bibitem{Mielkem0}A. Mielke, J. Phys. A{\bf 24},  L73  (1991).
\bibitem{Mielkem1}A. Mielke, J. Phys. A{\bf 24},  3311  (1991).
\bibitem{Mielkem2}A. Mielke, J. Phys. A{\bf 25},  4335  (1992).
\bibitem{Mielke99}A. Mielke, Phys. Rev. Lett. {\bf 82},  4212  (1999).
\bibitem{Tasakim1}H. Tasaki, Phys. Rev. Lett. {\bf 69},  1608  (1992).
\bibitem{mietasa}A. Mielke and H. Tasaki, Commun. Math. Phys. {\bf 158},
31 (1993).
\bibitem{Tasaki95}H. Tasaki, Phys. Rev. Lett. {\bf 75},  4678  (1995).
\bibitem{kohno}M. Kohno, Phys. Rev. B {\bf 56}, 15015 (1997).
\bibitem{Daul}S. Daul and R. Noack, Z. Phys. B {\bf 103},  293  (1997).
\bibitem{Daul2}S. Daul and R. Noack, Phys. Rev. B {\bf 58},  2635  (1997).
\bibitem{Liang}S. Liang and H. Pang, Europhys. Lett. {\bf 32}, 173 (1995).
\bibitem{saka}H. Sakamoto and K. Kubo, J. Phys. Soc. Jpn. {\bf 65}, 
3732 (1996).
\bibitem{Kusakabe94}K. Kusakabe and H. Aoki, Physica B
{\bf 194-196}, 217 (1994). 
\bibitem{Kusakabe}K. Kusakabe and H. Aoki, Phys. Rev. B {\bf 52},
R8684 (1995).
\bibitem{watamiya}Y. Watanabe and S. Miyashita,
J. Phys. Soc. Jpn. {\bf 66}, 3981 (1997).
\bibitem{Kubo}K. Kubo, J. Phys. Soc. Jpn. {\bf 51}, 782 (1982).
\bibitem{Sigrist}M. Sigrist, K. Ueda and H. Tsunetsugu,
Phys. Rev. Lett {\bf 67}, 2211 (1991).
\bibitem{KomaTasaki}T. Koma and H. Tasaki, J. Stat. Phys. {\bf 76}, 745 (1994).\bibitem{muller}E. M{\"u}ller-Hartmann, J. Low. Temp. Phys. {\bf 99}, 349 
(1995).
\bibitem{EderOhta}R. Eder and Y. Ohta, Phys. Rev. B {\bf 56}, R14247 (1997).
\bibitem{Arita}R. Arita, K. Kusakabe, K. Kuroki and H. Aoki,
Phys. Rev. B {\bf 58} R11833 (1998).
\bibitem{DoucotWen}B. Doucot and X.G. Wen, Phys. Rev. B 
{\bf 40}, 2719 (1989).
\end{references}
\end{document}